\documentclass[mathleft
]{an}
\usepackage{graphicx}
\usepackage{times}
\usepackage{amsmath}
\usepackage{flushend}
\usepackage{natbib}
\bibpunct{(}{)}{;}{a}{}{,}
\begin{document}

\Pagespan{954}{957}
\Yearpublication{2012}%
\Yearsubmission{2012}%
\Month{11}%
\Volume{333}%
\Issue{10}%
\DOI{10.1002/asna.201211830}%

\def\apj{ApJ}		
\def\aap{A\&A}		
\def\mnras{MNRAS}

\newcommand*{\secref}[1]{\S\ref{sec:#1}}

\title{On the possibility of using seismic probes to study the core composition in pulsating white dwarfs}

\author{J. P. Faria\inst{1,2}\fnmsep\thanks{
\email{Joao.Faria@astro.up.pt}\newline}
\and  M. J. P. F. G. Monteiro
\inst{1,2}
}
\titlerunning{On the core composition in pulsating white dwarfs}
\authorrunning{J.P.Faria \& M.J.P.F.P. Monteiro}
\institute{
Centro de Astrof\'{i}sica da Universidade do Porto, Rua das Estrelas, 4150-762 Porto, Portugal
\and 
Departamento de F\'{i}sica e Astronomia, Faculdade de Ci\^encias da Universidade do Porto
}

\received{2012 Nov 6}
\accepted{2012 Nov 7}
\publonline{later}

\keywords{stars: interiors -- stars: oscillations -- white dwarfs}

\abstract{%
White dwarfs correspond to the final stages of stellar evolution of solar-type stars. In these objects, production of energy by nuclear burning has ended which means that a white dwarf simply cools down over the course of the next billion years. It is now known that white dwarfs spend some of their cooling history in an instability strip. The pulsating white dwarfs with an hydrogen atmosphere (called DAV or ZZ Ceti stars) show non-radial oscillation modes with periods in the range 100 -- 1200s. In this work we try to illustrate how the oscillation \textit{p}-mode frequencies of idealized white dwarf models change as the result of a different chemical composition in the core, with the ultimate goal of determining the chemical stratification from seismic observations. The presence of acoustic glitches in the internal structure results in a periodic signal in the frequencies. We find that this signal depends on the chemical stratification/composition of the core in a form that can be analytically modelled.
}

\maketitle

\section{Introduction}
At the end of the main sequence phase, a small mass star has
exhausted the nuclear fuel in its core. It looses a significant part of the mass of its outer layers while in the AGB phase. Depending on when this mass loss ends (by dissipation of the rarefied expanded envelope) the carbon-oxygen core of the star will be surrounded with either a thin coating of hydrogen-rich material or a layer of nearly pure helium. We call the remaining star a white dwarf.

The hydrogen atmosphere (DA) white dwarfs comprise about 80\% of all spectroscopically known white dwarfs and are important, e.g., in understanding the mass loss process in the AGB phase \citep[e.g.][]{Marigo2012} and determining the age of the local galactic disk \citep{Wood1992}. Since the first detection of a pulsating white dwarf \citep{Landolt1968} the number of known pulsators has grown to about two hundred and it is now believed that these stars go through instability strips during the cooling process. The instabilities manifest in terms of multi-periodic luminosity variations which can be explained by non-radial \textit{g}-mode pulsations. This means that we can use the tools of asteroseismology to probe the internal structure of white dwarfs and determine some of their fundamental properties \citep[e.g.][]{Bradley1994, Corsico2012}. For a review on the recent state of seismology studies on white dwarfs see \citet{Winget2008} or \citet{Althaus2010}.

Despite never having been detected observationally \citep{Silvotti2011}, pressure-driven modes (\textit{p} modes) could also be excited in white dwarfs.
These are expected to show periods between $0.1$ and $10$~s \citep{Saio1983}. These modes probe the interior of the star differently as they are more sensitive to the structure of the core (\textit{g} modes are envelope modes). Our goal in this work is to study the effect that a different chemical composition in the core has on the oscillation frequencies of \textit{p} modes.

Very simple models of the structure of white dwarfs are computed using the appropriate equations (this is discussed in \secref{Equilibrium-Models}). The sharp transition caused by a chemical composition discontinuity gives rise to a characteristic signature in the oscillation frequencies. The nature of this signature, in the case of \textit{p} modes, is discussed in \secref{Effect} and in \secref{Results} results from the idealized models show the core composition dependence of the signal. The paper ends with a discussion of the results.

\section{\label{sec:Equilibrium-Models}Equilibrium Models}

\subsection{The Equation of State}

The pressure associated with matter at $T{=}0$ is what keeps the star from collapsing. The equation of state of this matter can be seen as that of an ideal, cold gas of fermions.  Following a usual convention we define the dimensionless Fermi momentum, usually called \emph{relativity parameter,} by $x{\equiv}\frac{p_{F}}{m_{e}c}$
so that we can write the pressure of the gas as \citep{Weiss2004}

\begin{equation}
\mathnormal{P=\frac{8\pi m_{e}^{4}c^{5}}{3h^{3}}\int_{0}^{x}\frac{x'^{4}}{\left(x'^{2}+1\right)^{1/2}}dx'=A\, f(x)} \, \label{eq:pressure}
\end{equation}

and the mass density

\begin{equation}
\rho=\frac{8\pi m_{e}^{3}c^{3}}{3h^{3}N_{A}}\,\mu_{e}\, x^{3}=B\,\mu_{e}x^{3}\,\textrm{,}\label{eq:density}
\end{equation}

where $A$ and $B$ are constants, $m_{e}$ is the mass of the electron, $h$ the Planck constant, $c$ the speed of light and $N_A$ the Avogadro number. The symbol $\mu_{e}$ represents the inverse of the number of electrons per unit atomic mass, sometimes refered to as the mean electron weight.

The essential role of the equation of state in this case is to specify the mechanical structure of the models, to which pulsation modes are very sensitive. Because the pressure in the interior of white dwarfs is completely dominated by the contribution of the cold and highly degenerate electron gas, we can expect this equation of state to be adequate in constructing models of pulsating white dwarfs. The goal here is not to develop a realistic model of the structure of white dwarfs but one that includes only the necessary physics to yield reliable frequency calculations.

\subsection{The Chemical Composition}
We have built several white dwarf models by integrating the hydrostatic equilibrium equation together with the equation of state defined by (\ref{eq:pressure}) and (\ref{eq:density}). The value of the central density is provided as input and iterated until the model has mass $M$. 

We consider an inner core of a specified chemical composition and mass surrounded by an envelope of pure hydrogen, thus neglecting the He layer that should be present in DA-type stars. The composition discontinuity in the core-envelope transition is handled by changing the value of the mean electron weight $\mu_{e}$. 

\begin{figure}
\centering
\includegraphics[width=\columnwidth]{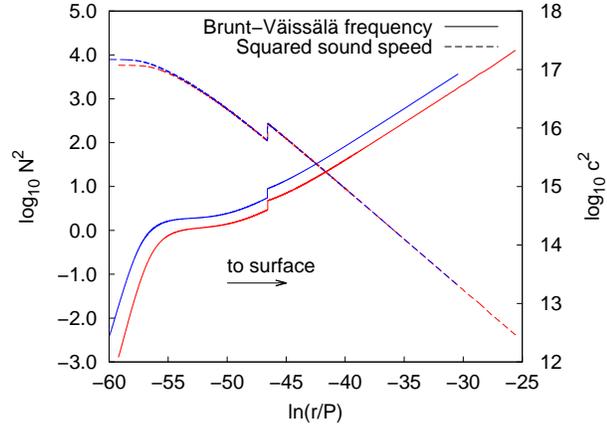}
\caption{The square of the Brunt-V\"{a}iss\"{a}l\"{a} frequency (solid lines) and the square of the sound speed (dashed lines) as a function of the natural logarithm of the ratio of radius to pressure. This coordinate is used to provide appropriate resolution at both the center and the surface of the model. The two colors represent models with different masses: $0.8\,\textrm{M}_{\odot}$
in red and $0.9\,\textrm{M}_{\odot}$ in blue. The two models have slightly different radii.}
\label{fig:N2andc2}
\end{figure}

The cores of typical white dwarfs are expected to be composed of the products of He burning: carbon and oxygen. There is however some observational and theoretical basis to expect up to iron-rich cores \citep[see][]{Hamada1961,Panei2000}. We define the following name convention for our models: model $M_{j}$ has a core composition of either $j$ = C (Carbon), O (Oxygen), I (Iron), M (Magnesium). We have also considered homogeneous Carbon models without the hydrogen layer, for comparison. 

\section{\label{sec:Effect}Effect on the frequencies}

The discontinuity in the chemical composition at the core-envelope
interface is expected to have an impact on the pulsation frequencies of the models, mainly due to its influence on the Brunt-V\"{a}iss\"{a}l\"{a} frequency $N$, which is defined as

\begin{equation}
N^{2}=g\left(\frac{1}{\Gamma_{1}}\frac{d\ln P}{dr}-\frac{d\ln\rho}{dr}\right)\textrm{,}
\end{equation}

where $g{=}Gm/r^{2}$ is the gravitational acceleration, $G$ the
gravitational constant and $\Gamma_{1}{\equiv}(d\ln P/d\ln\rho)_s$ is the first adiabatic exponent (derivative at constant entropy $s$). This is the natural characteristic frequency that dominates the calculation of pulsation periods in white dwarfs. Any sharp feature (in our case a discontinuity) on the run of $N$ as a function of depth (see Figure~\ref{fig:N2andc2}) will have an effect on the frequencies. In this section we briefly describe a method used in solar-type stars that allows us to predict what this effect will be, in the case of \textit{p} modes. 

Following \citep{Christensen-Dalsgaard1995}, let $V(\tau)$ be the \emph{acoustic potential} which can be written as 

\begin{equation}
\begin{split}
V^{2} &= N^{2} + \frac{c^{2}}{4}\left(\frac{2}{r}{+}\frac{N^{2}}{g}{-}\frac{g}{c^{2}}{-} \frac{1}{2c^{2}}\frac{dc^{2}}{dr} \right)^{2} \\
& {-}\frac{c}{2}\frac{d}{dr}\left[c\left(\frac{2}{r}+\frac{N^2}{g}-\frac{g}{c^2}-\frac{1}{2c^2}\frac{dc^2}{dr}\right)\right] 
\!{-}4\pi G\rho \,\textrm{,}
\label{eq:potential}
\end{split}
\end{equation}

where $c$ is the sound speed and $\tau$ is the acoustic depth defined by

\begin{equation}
\tau(r)=\int_{r}^{R}\frac{dr'}{c}\,\textrm{.}
\end{equation}

The acoustic size of the star is $\tau_s=\tau(0)$, while $R$ is the radius of the star. A discontinuity in the sound speed (Figure \ref{fig:N2andc2}) due to the chemical stratification gives rise to a discontinuity in the potential $V(\tau)$. 

The eigenfunctions $Y$ of the modes of oscillation satisfy, asymptotically and for low degree $\ell$, the equation \citep[e.g.][]{Vorontsov1989}

\begin{equation}
\frac{d^{2}Y}{d\tau^{2}}+\left[\omega^{2}-V^{2}(\tau)\right]Y=0 \,\textrm{,}\label{eq:asymptoEq}
\end{equation}

where $\omega$ is the frequency of the mode. For the case of a potential with a step function of amplitude $\delta V$, we can solve equation (\ref{eq:asymptoEq}) and show that there will be a periodic signal arising in the frequencies \citep[see][]{Monteiro1994}. The change in the the frequencies, $\delta\omega$, relative to the unperturbed frequency $\omega_{0}$, will be 

\begin{equation}
\delta\omega\sim\frac{\delta V^{2}}{4\tau_{s}\omega_{0}^{2}}\sin\left(2\omega_{0}\tau_{d}\right)=A_{0}\left(\frac{\omega_{r}}{\omega_{0}}\right)^{2}\sin\left(2\omega_{0}\tau_{d}\right)\,\textrm{,}
\end{equation}

\begin{figure}
\includegraphics[width=\columnwidth]{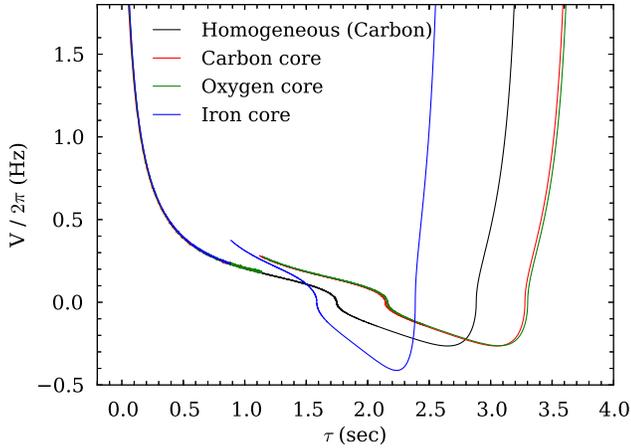}
\caption{Acoustic potential as a function
of the acoustic depth $\tau$ (note that $\tau=0$ is the surface).
The four models shown have $0.9\textrm{M}_{\odot}$ and different
core compositions. The black line corresponds to the homogeneous carbon
model, without the hydrogen layer. The step in the potential has a
different amplitude and location for each model. }
\label{fig:potential}
\end{figure}

with $\tau_{d}$ the acoustic depth at which the discontinuity occurs. We use a reference frequency $\omega_{r}/2\pi=10\,$Hz (representative value within the frequency range for $p$~modes).

Note that both the amplitude (which is a decreasing function of frequency) and the periodicity of the signal are completely determined by the amplitude and location of the step in the potential. In Figure (\ref{fig:potential}), $V(\tau)$ is plotted for four different models, each with a different core composition. For the homogeneous model there is no step function and thus we expect no periodic signal. 

\section{\label{sec:Results}Results}

\begin{figure}
\centering
\includegraphics[width=\columnwidth]{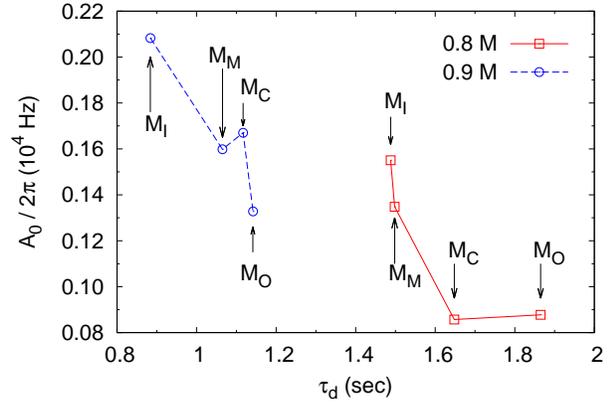}
\caption{Results for the amplitudes and acoustic locations
of the discontinuity for models with $0.8\textrm{M}_{\odot}$ (red)
and $0.9\textrm{M}_{\odot}$ (blue). Each point represents a different core composition, as indicated by the labels. }
\label{fig:amplitudes}
\end{figure}

For models with $0.8\,\textrm{M}_{\odot}$ and $0.9\,\textrm{M}_{\odot}$, the results for the amplitude $A_{0}$ and the acoustic depth $\tau_{s}$ of the discontinuity on the potential are shown in Figure (\ref{fig:amplitudes}). Each different composition results in a different point in this graph, meaning that the signal in the frequencies will depend on the core composition of the model. 

\begin{figure}
\centering
\includegraphics[width=\columnwidth]{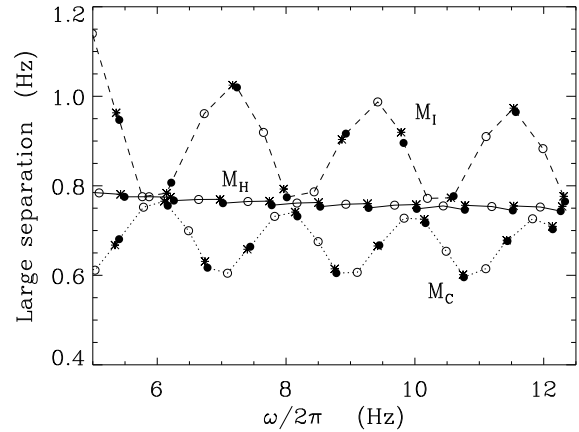}
\caption{Large separation as a function of frequency for
three representative models of $0.8\textrm{M}_{\odot}$ with an homogeneous
carbon composition (solid line), a carbon core (dotted line) and an
iron core (dashed line). Filled circles represent $\ell=0$ modes,
open circles $\ell=1$ and stars $\ell=2$.}
\label{fig:signal}
\end{figure}

Using the POSC code \citep{Monteiro2008}, linear adiabatic frequencies were calculated for each model, for modes with $\ell=0,1$ and $2$. The large separations, defined as the difference between the frequencies of consecutive modes of the same degree $\ell$, were also computed. The predicted periodic signal appears on this quantity, as is visible on Figure (\ref{fig:signal}), for three representative models of $0.8\textrm{M}_{\odot}$. The trend in the amplitude of the signal with the frequency of the mode is noticeable in the $M_{I}$ model.
The periodicity is also different for the two models with a discontinuity.

\section{\label{sec:Discussion}Discussion}

Very simple equilibrium models of white dwarf stars were built in order to calculate oscillation frequencies. The hydrostatic equilibrium equation was integrated, considering the equation of state of a degenerate, ideal gas of electrons. It is clear that this part of our work requires a more adequate approach and that a study with more realistic models (full evolutionary models that take into account the cooling process and a more realistic equation of state) is required. 

The equation of state is not the major problem though: the treatment of the chemical stratification done here is simplistic. However, our models contain the relevant aspects of the physics of cool white dwarfs that may be probed with seismic data and are, thus, capable of providing useful results.

Calculating the size and location of the step function in the acoustic potential $V(\tau)$ allowed us to predict the effect that the core-envelope discontinuity has on the \textit{p}-mode frequencies: a periodic signal. This signal is clearly seen in the model frequencies (Figure \ref{fig:signal}) and depends substantially on the chemical composition of the core. The absence of any periodic variation in the homogeneous model also confirms our predictions. The periodic signal in the large frequency separation displays a variation of amplitude with frequency and a periodicity associated with the acoustic location of the discontinuity, as predicted. These results open the window for further detailed analysis of both more accurate models and observational seismic data of white dwarfs.

The asymptotic analysis considered in \secref{Effect} is valid for \textit{p}-mode frequencies and is commonly used in the context of helioseismology and asteroseismology of solar-type stars (see e.g., \citealt{Monteiro2000}; \citealt{Mazumdar2012}).

We emphasize again that, until now, only \textit{g}-mode oscillation frequencies have been detected in white dwarfs. The very high frequencies expected for \textit{p} modes make these very hard to detect with present day instruments, even if they are indeed excited. This does not invalidate our analysis as the sharp features also give rise to a similar effect in the period spacings of \textit{g} modes \citep{Metcalfe2001}.

\acknowledgements Part of this work was supported by FCT-MEC-Portugal and FEDER-EC, through grant PTDC/CTE-AST/098754/2008.



\begin{thebibliography}{19}
\expandafter\ifx\csname natexlab\endcsname\relax\def\natexlab#1{#1}\fi

\bibitem[{{Althaus} {et~al.}(2010){Althaus}, {C{\'o}rsico}, {I}sern, \&
  {Garc{\'{\i}}a-Berro}}]{Althaus2010}
{Althaus}, L.~G., {C{\'o}rsico}, A.~H., {I}sern, J.,  {Garc{\'{\i}}a-Berro},
  E.: 2010, A\&ARv~18, 471

\bibitem[{{Bradley} \& {Winget}(1994)}]{Bradley1994}
{Bradley}, P.~A., {Winget}, D.~E.: 1994, ApJ~430, 850

\bibitem[{{Christensen-Dalsgaard} {et~al.}(1995){Christensen-Dalsgaard},
  {Monteiro}, \& {Thompson}}]{Christensen-Dalsgaard1995}
{Christensen-Dalsgaard}, J., {Monteiro}, M.~J.~P.~F.~G., {Thompson}, M.~J.:
  1995, MNRAS~276, 283

\bibitem[{{C{\'o}rsico} {et~al.}(2012){C{\'o}rsico}, {Althaus}, {Miller
  Bertolami}, \& {Bischoff-Kim}}]{Corsico2012}
{C{\'o}rsico}, A.~H., {Althaus}, L.~G., {Miller Bertolami}, M.~M., 
  {Bischoff-Kim}, A.: 2012, A\&A~541, A42

\bibitem[{{Hamada} \& {Salpeter}(1961)}]{Hamada1961}
{Hamada}, T., {Salpeter}, E.~E.: 1961, ApJ~134, 683

\bibitem[{{Landolt}(1968)}]{Landolt1968}
{Landolt}, A.~U.: 1968, ApJ~153, 151

\bibitem[{{Marigo}(2012)}]{Marigo2012}
{Marigo}, P.: 2012, in IAU Symposium, Vol. 283, IAU Symposium, 87--94

\bibitem[{{Mazumdar} {et~al.}(2012){Mazumdar}, {Michel}, {Antia}, \&
  {Deheuvels}}]{Mazumdar2012}
{Mazumdar}, A., {Michel}, E., {Antia}, H.~M., {Deheuvels}, S.: 2012, A\&A~540, A31

\bibitem[{{Metcalfe} {et~al.}(2001){Metcalfe}, {Winget}, \&
  {Charbonneau}}]{Metcalfe2001}
{Metcalfe}, T.~S., {Winget}, D.~E., {Charbonneau}, P.: 2001, ApJ~557, 1021

\bibitem[{{Monteiro}(2008)}]{Monteiro2008}
{Monteiro}, M.~J.~P.~F.~G.: 2008, Ap\&SS~316, 121

\bibitem[{{Monteiro} {et~al.}(1994){Monteiro}, {Christensen-Dalsgaard}, \&
  {Thompson}}]{Monteiro1994}
{Monteiro}, M.~J.~P.~F.~G., {Christensen-Dalsgaard}, J., {Thompson}, M.~J.:
  1994, A\&A~283, 247

\bibitem[{{Monteiro} {et~al.}(2000){Monteiro}, {Christensen-Dalsgaard}, \&
  {Thompson}}]{Monteiro2000}
{Monteiro}, M.~J.~P.~F.~G., {Christensen-Dalsgaard}, J., {Thompson}, M.~J.:
  2000, MNRAS~316, 165

\bibitem[{{Panei} {et~al.}(2000){Panei}, {Althaus}, \& {Benvenuto}}]{Panei2000}
{Panei}, J.~A., {Althaus}, L.~G., {Benvenuto}, O.~G.: 2000, MNRAS~312, 531

\bibitem[{{Saio} {et~al.}(1983){Saio}, {Winget}, \& {Robinson}}]{Saio1983}
{Saio}, H., {Winget}, D.~E., {Robinson}, E.~L.: 1983, ApJ~265, 982

\bibitem[{{Silvotti} {et~al.}(2011){Silvotti}, {Fontaine}, {Pavlov}, {Marsh},
  {Dhillon}, {Littlefair}, \& {Getman}}]{Silvotti2011}
{Silvotti}, R., {Fontaine}, G., {Pavlov}, M., {et~al.}: 2011, A\&A~525, A64

\bibitem[{{Vorontsov}, {Zharkov}(1989)}]{Vorontsov1989}
{Vorontsov}, S.~V., {Zharkov}, V.~N.: 1989, ASPRv~7, 1

\bibitem[{Weiss {et~al.}(2004)Weiss, Hillebrandt, Thomas, \&
  Ritter}]{Weiss2004}
Weiss, A., Hillebrandt, W., Thomas, H.-C., Ritter, H.: 2004, {C}ox and
  {G}uili's {P}rinciples of {S}tellar {S}tructure, 2nd edn., Advances in
  Astronomy \& Astrophysics (Cambridge Scientific Publishers)

\bibitem[{{Winget} \& {Kepler}(2008)}]{Winget2008}
{Winget}, D.~E., {Kepler}, S.~O.: 2008, ARA\&A~46, 157

\bibitem[{{Wood}(1992)}]{Wood1992}
{Wood}, M.~A.: 1992, ApJ~386, 539

\end{thebibliography}
\end{document}